\documentclass[aps,prb,reprint,showpacs,superscriptaddress,longbibliography]{revtex4-2} %
\usepackage{mathtools,amssymb,graphicx,units}
\usepackage[usenames,dvipsnames]{color}
\usepackage[plainpages=false,pdfpagelabels,colorlinks=true,linkcolor=red,urlcolor=blue,citecolor=blue,pdftitle={Title},pdfauthor={},pdfdisplaydoctitle=true,pdfduplex=DuplexFlipLongEdge]{hyperref}
\usepackage{verbatim}
\usepackage[english]{babel}

\usepackage[normalem]{ulem}
\usepackage{appendix}
\usepackage{bbm}
\usepackage{makecell}

\newcommand{\beginsupplement}{%
        \setcounter{table}{0}
        \renewcommand{\thetable}{S\arabic{table}}%
        \setcounter{figure}{0}
        \renewcommand{\thefigure}{S\arabic{figure}}%
}
\graphicspath{{./fig/}}

\begin{document}
\title{Bound states in one-dimensional systems with colored noise}
\author{Xingbo Wei}
\email{weixingbo@zstu.edu.cn}
\affiliation{Zhejiang Key Laboratory of Quantum State Control and Optical Field Manipulation, Department of Physics, Zhejiang Sci-Tech University, 310018 Hangzhou, China}
\author{Kewei Feng}
\affiliation{Zhejiang Key Laboratory of Quantum State Control and Optical Field Manipulation, Department of Physics, Zhejiang Sci-Tech University, 310018 Hangzhou, China}
\author{Tian-cheng Yi}
\affiliation{Zhejiang Key Laboratory of Quantum State Control and Optical Field Manipulation, Department of Physics, Zhejiang Sci-Tech University, 310018 Hangzhou, China}
\author{Tong Liu}
\affiliation{Department of Applied Physics, School of Science, Nanjing University of Posts and Telecommunications, Nanjing 210003, China}

\author{Gao Xianlong}
\affiliation{Department of Physics, Zhejiang Normal University, Jinhua 321004, China}

\author{Yunbo Zhang}
\email{ybzhang@zstu.edu.cn}
\affiliation{Zhejiang Key Laboratory of Quantum State Control and Optical Field Manipulation, Department of Physics, Zhejiang Sci-Tech University, 310018 Hangzhou, China}

\begin{abstract}
We investigate the phase transitions in a one-dimensional system with colored noise. Previous studies indicated that the phase diagram of this system included extended and disorder-induced localized phases. However, by studying the properties of wave functions, we find that this phase diagram can be further refined, revealing the existence of a bound phase for the large potential amplitude $W$ and noise control parameter $\alpha$. In the bound phase, the wave function cannot extend throughout the entire chain, tails decay faster than exponentially and its distribution expands as the system size increases. By adjusting the potential amplitude to induce a transition from the extended phase to the bound phase, we find that bound states coexist with extended states in the spectrum. In contrast, when the system transitions from the Anderson localized phase to the bound phase, we do not observe the obvious coexistence of Anderson localized and bound states. Finally, by performing the time evolution, we find that the dynamic transition point of the bound phase is inconsistent with the static one for large $\alpha$.
\end{abstract}

\maketitle
\section{Introduction}
White noise is commonly used in studies of Anderson localization, where it causes the wave function to decay exponentially in space~\cite{AndersonAbsence}. Besides white noise, various other types of colored noise exist in nature, such as pink noise, blue noise, etc. The ``color" of noise refers to its power spectrum, and different colored noise can be characterized by the power spectral density $S(k) = \int_{-\infty}^{\infty} R(x) e^{-i 2 \pi k x} \, dx$~\cite{zhivomirovmethod,PChoi022122,Tessieri066120,Cohen2292}, where $R(x)=\langle W_j W_{j+x}\rangle$ is the autocorrelation function of colored noise between sites $j$ and $j+x$. Different types of colored noise exhibit distinct decay forms as $S(k) \propto k^{-\alpha}$. Typically, white noise has a flat spectral density with $\alpha = 0$, pink noise has $\alpha = 1$, and red noise has $\alpha = 2$~\cite{zhivomirovmethod}. An important property of colored noise is that the correlation changes with the exponent $\alpha$~\cite{MouraDisorder,ShimaLocal,MouraDeloc}. For $\alpha = 0$, the white noise manifests $\delta$-correlation $\langle W_j W_{j'}\rangle=\langle W_j^2\rangle\delta_{j,j'}$ in space~\cite{MouraDisorder}, which causes one-dimensional systems to form Anderson localized states for arbitrary potential amplitude~\cite{Abrahams}. As $\alpha$ increases, the correlation of colored noise alters from short-range to long-range, and a transition from the Anderson localized phase to the extended phase is found in the middle of the spectrum~\cite{ShimaLocal,MouraDisorder,MouraDeloc,izrailev2012anomalous,kaya2007localization,croy2011anderson,khan2023anomalous,Santos205148,Moura174203,Lima104416}. 

\begin{figure}[htbp]
	\includegraphics[width=1.0\columnwidth,height=0.8\columnwidth]{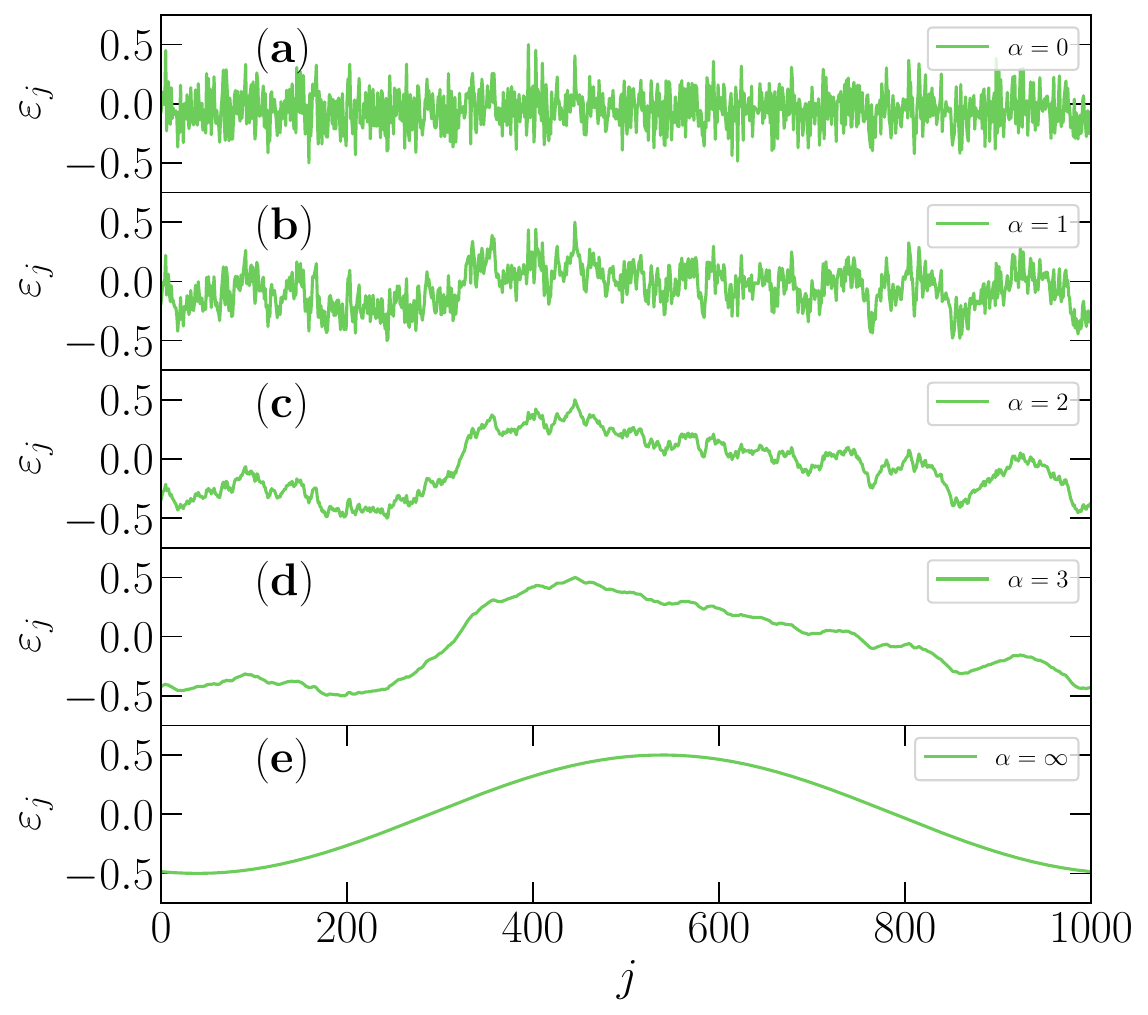}
	\vspace{-0.4cm}
	\caption{The noise signals in space for different $\alpha$. (a), (b), (c), and (d) use the same random phase $\phi_k$. The amplitude interval is normalized to $\varepsilon_j\in[-1/2, 1/2]$ and $L=1000$. 
	}
	\label{fig1}
\end{figure}

In the studies of Anderson localization, the localization length $\xi$ is a crucial quantity, which identifies the exponential decay rate of the wave function $\psi_r\sim e^{-r/\xi}$~\cite{Leesystems}. The localization length of Anderson localized states can be calculated by the Lyapunov exponent $\gamma$ in many systems with white noise or quasi-periodic potentials~\cite{ZhouExact,Zhang174206}. This is due to the fact that the Lyapunov exponent is equivalent to the reciprocal of the localization length $\gamma=1/\xi$ when a wave function obeys the exponential decay $\psi_r\sim e^{-\gamma r}$~\cite{Zhang174206}. On the contrary, when the wave function does not follow exponential (Anderson-like) decay, the above relation $\gamma=1/\xi$ is not strictly valid. Therefore, the localization length derived from the Lyapunov exponent may not accurately characterize states with non-exponential (non-Anderson-like) decay, e.g., Wannier-Stark states~\cite{Longhi}. 

In previous works, the Lyapunov exponent is used to characterize the transition in one-dimensional systems with colored noise or long-range correlated disorder~\cite{ShimaLocal,MouraDisorder,MouraDeloc,izrailev2012anomalous,kaya2007localization,zhao2010critical, PhysRevB.72.174207}. An important phase diagram shows that the system is in the localized phase for the large potential amplitude $W/t>4$ and $0 \leq\alpha\leq 5$~\cite{ShimaLocal}. However, one can find that the randomness of the noise signal decreases as $\alpha$ increases and the signal becomes smooth in the limit for $\alpha\rightarrow \infty$, as shown in Fig.~\ref{fig1}. It raises an important question: are the states still Anderson localized states for large potential amplitude $W/t>4$ and large $\alpha$? In other words, could there be a phase different from the Anderson localized phase?  In this work, our objective is to address this problem. By carefully studying the wave function and its characteristics, we confirm that a bound phase exists for large $W$ and $\alpha$, in which the wave function does not obey exponential (Anderson-like) decay, suggesting that the Lyapunov exponent cannot accurately describe the properties of the system with long-range correlated disorder. Moreover, we also investigate the coexistence of different states in the spectrum and perform the time evolution to study the dynamical properties of the bound states for large $\alpha$.

\section{Model}
We study a single atom in a one-dimensional chain with colored noise under periodic boundary conditions. The Schr\"odinger equation is expressed as
\begin{equation}\label{hub}
t\left(\psi_{j+1}+\psi_{j-1}\right)+W\varepsilon_j\psi_j=E \psi_j,
\end{equation}
where $\psi_j$ denotes the amplitude of the wave function at site $j$. $t\equiv1$ is the nearest-neighbor hopping energy, $W$ is the amplitude of the potential, and $E$ is the eigenvalue. $\varepsilon_j=-1/2+(\tilde{\varepsilon}_j-\min\{\tilde{\varepsilon}_j\})/(\max\{\tilde{\varepsilon}_j\}-\min\{\tilde{\varepsilon}_j\})$ represents the normalized colored noise, modulating the interval to $[-1/2, 1/2]$. The non-normalized colored noise $\tilde{\varepsilon}_j$ is defined as~\cite{MouraDisorder,nguezAdame}
\begin{equation}\label{potential}
	\tilde{\varepsilon}_j=\sum_{k=1}^{L / 2} k^{-\alpha/2} \cos \left(\frac{2 \pi j k}{L}+\phi_k\right),
\end{equation}
where $L$ is the number of sites and $\phi_k$ are $L/2$ $k$-dependent random phase factors, which are uniformly distributed within the interval $[0, 2\pi)$. Eq. \eqref{potential} is widely used to study the localization transition induced by long-range correlated disorder~\cite{MouraDisorder,MouraDeloc,izrailev2012anomalous,kaya2007localization}. Additionally, the Fourier filtering method can also generate long-range correlated disorder~\cite{ShimaLocal}. Through careful inspection, we determine that both methods yield consistent conclusions regarding the bound phase. Here, we choose Eq. \eqref{potential} as the object because it has a clear expression, which is conducive to theoretical analysis. We regulate $\tilde{\varepsilon}_j$ to $\varepsilon_j$ by using the above scaling formula. The main reason for this adjustment is that we find the bound effect of the potential is more dominant than the disorder effect for large $\alpha$. Note that the essence of the bound effect is the competition between the hopping energy and the potential amplitude, thus the transition is determined by the potential amplitude for a given hopping energy~\cite{Wei134207,Wei023314}. Previous studies have regulated either the potential amplitude~\cite{ShimaLocal,zhao2010critical} or the variance of the potential~\cite{MouraDisorder, PhysRevLettKantelhardt, PhysRevLett.84.199,zhang2002localization,kaya2007localization,khan2019spectral,khan2023entanglement,assunccao2011coherent,nguyen2011influence}. Based on the findings in this work, we argue that regulating the potential amplitude provides a more accurate reflection on the phase transition for large $\alpha$. For comparison, we also provide the results in the case of regulating the variance of the potential in Appendix~\ref{add3}.

\begin{figure}[htbp]
	\includegraphics[width=1.0\columnwidth,height=0.8\columnwidth]{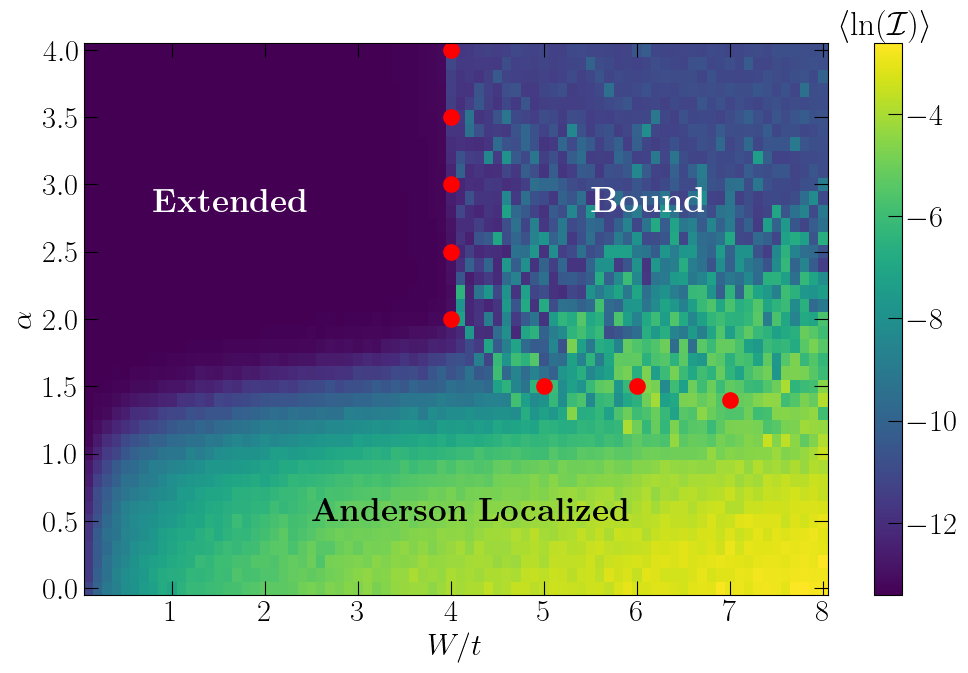}
	\vspace{-0.4cm}
	\caption{Phase diagram in the middle of the spectrum $E_{target}\approx0$. The colors represent the values of $\ln(\langle\mathcal{I}\rangle)$, which are obtained from Eq. \eqref{hub} of size $L=10^6$, using 20 samples. The red marks represent the transition points indicated by the fractal dimension between the bound phase and other phases.
	}
	\label{fig2}
\end{figure}

To characterize different phases, we use multiple quantities in this work. The first one is the wave function, whose spatial distribution and the decay of the tail provide an intuitive distinction between different phases. The second quantity is the inverse participation ratio (IPR) $\mathcal{I}=\sum_j|\psi_j|^{4}$~\cite{Li064203}. For localized states, $\mathcal{I}$ is size-independent, whereas $\mathcal{I}$ decays as the size increases for non-localized states~\cite{Liu104201}. Furthermore, we also use IPR-related fractal dimensions $D=-\log(\mathcal{I})/\log(L)$ to characterize the localization transition. For localized (extended) states $D=0$ ($D=1$), whereas $0<D<1$ for multifractal states~\cite{Duthie}.   

In this work, we introduce the state-of-the-art shift-invert algorithm~\cite{pietracaprina2018shift} to study the transition between different phases in the middle of the spectrum $E_{target}\approx0$, and vary the target energy $E_{target}$ to investigate the coexistence of different states. To investigate the dynamical properties, we use the exact diagonalization algorithm to perform the time evolution.

\section{Results}
\subsection{the transition at $E_{target}\approx0$}
In Fig.~\ref{fig2}, we map out the phase diagram in the middle of the spectrum at $E_{target}\approx0$, which encloses an extended phase, an Anderson localized phase, and a bound phase. Different from previous work where the phase diagram only consists of extended and localized phases~\cite{ShimaLocal}, in our work we confirm that a new phase named the bound phase exists for large $\alpha$ and $W$. The characteristics of three quantum phases are shown in Table~\ref{table0}. In the extended phase, the wave function is distributed in the whole space without decay and the fractal dimension $D=1$ for $L\rightarrow \infty$~\cite{Duthie}. On the contrary, the wave function in the Anderson localized phase is only distributed in partial space, whose tail decays exponentially as $\psi_r\propto e^{-r/\xi}$~\cite{AndersonAbsence,Leesystems}. Furthermore, one can also find that the fractal dimension $D=0$ for $L\rightarrow \infty$ in the Anderson localized phase~\cite{Duthie}. Different from the previous two phases, the bound phase shows exotic properties, which contains part characteristics of the Anderson localized states and extended states. Specifically, the wave function of the bound phase is not distributed in the whole space. This characteristic is similar to the Anderson localized states. However, by carefully studying the decay of the wave function, one can find that the tails of the bound states have higher than exponential decay, i.e., 
$-\ln\psi_{r}\propto r^{b}$ with $b>1$~\cite{Longhi,Wei023314,zhao2024fatepseudomobilityedgemultistates}, where $r$ is distance. This is different from the exponential decay of Anderson localized states, $-\ln\psi_{r}\propto r$. More importantly, the wave function in the bound phase expands as the size increases and the fractal dimension is $D=1$ for $L\rightarrow \infty$. This feature is nonetheless the property of an extended state~\cite{PhysWang,PhysRevB.96Xiao,dos2007critical}. In some works, these bound states are also referred to weakly ergodic states because they are extended with system sizes but not ergodic~\cite{Wei023314,acker032117,zhao2024fatepseudomobilityedgemultistates}.

\begin{figure}[htbp]
	\includegraphics[width=1.0\columnwidth,height=0.8\columnwidth]{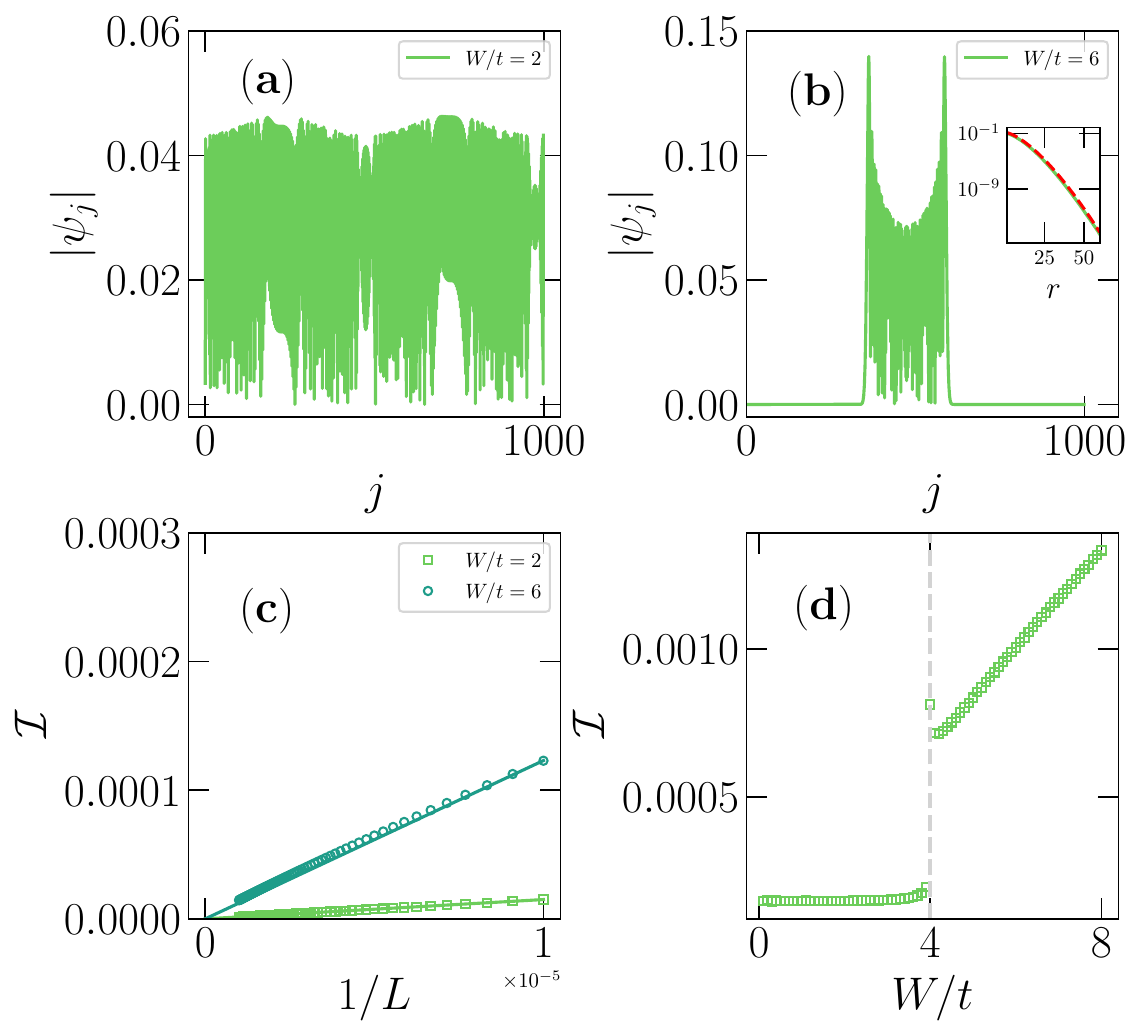}
	\vspace{-0.4cm}
	\caption{(a) and (b) Typical wave functions of extended states and bound states. (c) The inverse participation rate $\mathcal{I}$ as a function of $1/L$. (d) The inverse participation rate $\mathcal{I}$ as a function of $W/t$. $\tilde{\varepsilon}_j = \cos(2\pi j/L+\phi_{1})$ for $\alpha=\infty$ and $\phi_1=\sqrt{3}$. The inset in (b) indicates the decay of one tail of the wave function and the fitting function is $|\psi_{588+r}|=\exp[-0.0783*r^{1.477}-1.89]$ (the red dashed line). The grey dashed line in (d) marks the transition point $W/t=4$. In (d), we use $L=10000$.
	}
	\label{fig3}
\end{figure}

\begin{table}[htbp]
     \centering
	\caption{The characteristics of the extended phase, the Anderson localized phase, and the bound phase.}
	\begin{tabular}{|c|c|c|c|c}
		\cline{1-4}
		& \thead{Extended phase } &\thead{ Anderson \\ Localized phase} & \thead{Bound phase}                &  \\ 
		\cline{1-4}
		\thead{Distribution of \\ the wave function} & \thead{whole space}         & \thead{partial space}         & \thead{partial space}                 &  \\ 
		\cline{1-4}
		\thead{Decay of the tail}                 & \thead{no decay}            & \thead{exponential \\ decay}     & \thead{higher than \\ exponential \\ decay} &  \\ 
		\cline{1-4}
		\thead{Fractal dimension}       & $D=1$       & $D=0$      & $D=1$                 &  \\
		\cline{1-4}
	\end{tabular}
	
	\label{table0}
\end{table}

To understand the phase diagram phenomenologically, we first consider two limiting cases of $\alpha=0$ and $\alpha=\infty$. For $\alpha=0$, the potential behaves as white noise, it is well-known that the system is in the Anderson localized phase regardless of the strength of the random potential in one-dimensional systems~\cite{Abrahams}. To illustrate another limit, we rewrite Eq. \eqref{potential} as
\begin{equation}\label{potential2}
	\begin{aligned}
	\tilde{\varepsilon}_j=&\cos \left(\frac{2 \pi j}{L}+\phi_1\right)\\
	&+\sum_{k=2}^{L / 2}k^{-\alpha/2} \cos \left(\frac{2 \pi j k}{L}+\phi_k\right).
    \end{aligned}
\end{equation}
For $\alpha=\infty$,  the first term in Eq. \eqref{potential2} is much larger than the others, thus the potential can be simplified to $\tilde{\varepsilon}_j= \cos(2\pi j/L+\phi_{1})$, which is an ordered potential, as shown in Fig.~\ref{fig1} (e). 
The order of the potential is the prime factor in the formation of bound states~\cite{diaz2005anomalous}.
In this limit, we verify that the system transitions from the extended phase to the bound phase as the amplitude of the potential increases. 
In Fig.~\ref{fig3} (a) and (b), we show the typical extended state for $W/t=2$ and the bound state for $W/t=6$. One can find that the extended state is distributed throughout the entire chain, whereas the bound state is confined to part of the space. It needs to be pointed out that the potential is ordered for $\alpha=\infty$, thus there is no disorder effect. The formation of the bound phase is entirely due to the bound effect of the potential. Different from the Anderson localized states whose tail decays exponentially~\cite{Leesystems}, the tail of the bound state has a higher than exponential decay. In the inset of Fig.~\ref{fig3} (b), we use $|\psi_{j_0+r}|=\exp[-ar^{b}+c]$ to fit the tails, where $j_0$ is the starting position and $r$ represents the distance. $a$, $b$ and $c$ denote the fitting coefficients. The fitting function shows that the wave function decays as $|\psi_{588+r}|=\exp[-0.0783r^{1.477}-1.89]$, where $b=1.477>1$ indicates that the tail of the wave function exhibits faster than exponential decay. Ref.~\cite{Wei023314} shows that the Lyapunov exponent is $\gamma>0$ for a wave function with higher than exponential decay, thus the bound states may be misjudged as localized states if they are only characterized by the Lyapunov exponent. 
Furthermore, it should be noted that the bound state has double-peaked structure~\cite{diaz2005anomalous}, which contrasts with the single-peak structure observed in the Anderson localized states.
In Fig.~\ref{fig3} (c), IPRs of both extended and bound states decrease with $1/L$, suggesting that their wave functions expand as the system size increases. In the thermodynamic limit, the fractal dimension of the bound state is $D=1$. Therefore, strictly speaking, bound states cannot be considered as localized states because their wave functions scale with the system size. To find the transition point between the extended phase and the bound phase, we study the IPR as a function of $W/t$ in Fig.~\ref{fig3} (d), where the IPR shows a dramatic change at $W/t=4$. This transition point can be captured from the perspective of energy conservation. In Eq. \eqref{hub}, the hopping term has the energy bandwidth $4t$. An ordered potential needs the amplitude $W \geq 4t$ to bind the atom. Thus, it can be concluded that the transition point is at $W/t=4$.

\begin{figure}[htbp]
	\includegraphics[width=1.0\columnwidth,height=0.8\columnwidth]{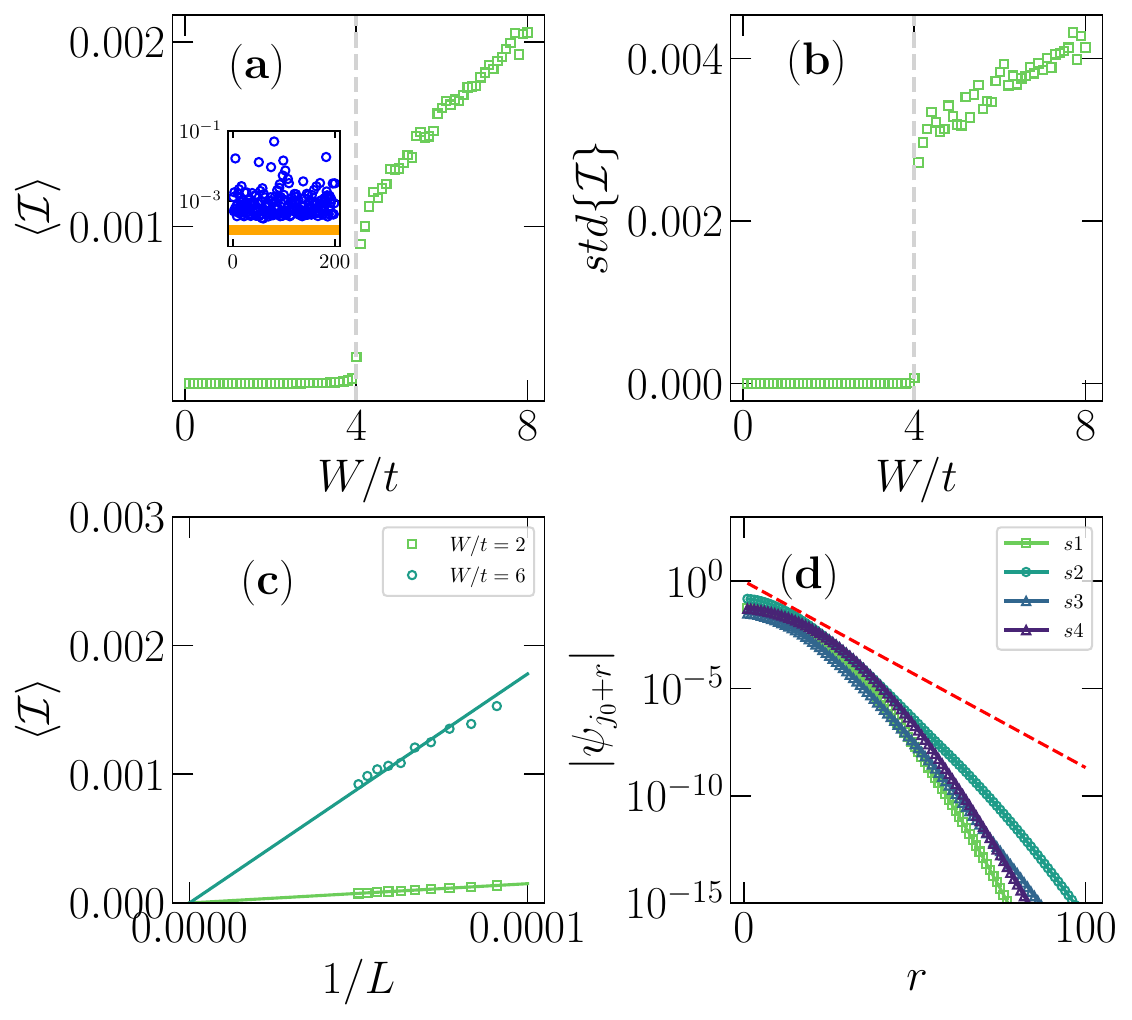}
	\vspace{-0.4cm}
	\caption{(a) The mean IPR as a function of $W/t$. (b) The standard deviation of IPR as a function of $W/t$. (c) The scaling of the mean IPR for different $W/t$. (d) The tails of bound states for $W/t=6$ (four samples). The noise control parameter is $\alpha=3$. We use 10000 samples in (a), (b) and (c).  $L=10000$ and the grey dotted lines in (a), (b) mark the transition point $W/t=4$. The inset in (a) shows partial IPR samples for $W/t=2$ (orange square) and $W/t=6$ (blue circle). In (d), the red dashed line represents the exponential decay with $b=1$, and the fitting exponents $b$ of four samples are: 1.77, 1.40, 1.45, 1.82, respectively.
	}
	\label{fig4}
\end{figure}

Next, we choose $\alpha=3$ to study the case away from these two limits. For $\alpha=3$, different $\phi_k$ lead to different potential distributions. Therefore, we use a large number of samples and take the average. In Fig.~\ref{fig4} (a), we show the mean inverse participation ratio $\langle \mathcal{I} \rangle$ as a function of $W/t$, where $\langle \mathcal{I} \rangle$ alters dramatically at the transition point $W/t=4$, which is the same as the case of $\alpha=\infty$. To further demonstrate the properties of different samples, we show the $\mathcal{I}$ for different samples in the inset of Fig.~\ref{fig4} (a). One can find that different samples have similar $\mathcal{I}$ in the extended phase, whereas $\mathcal{I}$ is highly dependent on the samples in the bound phase. It causes the standard deviation of $\mathcal{I}$ in the bound phase to be significantly larger than that in the extended phase, as shown in Fig.~\ref{fig4} (b). By carefully studying the distribution of wave functions, we confirm that the larger fluctuation of $\mathcal{I}$ in the bound phase is due to the distribution of wave functions being highly dependent on the potential distribution~\cite{Nishino033105}. In Fig.~\ref{fig4} (c), we study the scaling of $\langle \mathcal{I} \rangle$, in which $\langle \mathcal{I} \rangle$ of both the extended and bound states decreases with $1/L$, suggesting that the wave functions expand with the increasing size. In Fig.~\ref{fig4} (d), we show the tails of bound states. Here we display four samples. Although $\alpha=3$ is far away from the limit $\alpha=\infty$, the tails decay smoothly with increasing distance $r$, similar to the behavior observed for $\alpha=\infty$ in Fig.~\ref{fig3} (b). To further study the decay of the tails, we also use the function $|\psi_{j_0+r}|=\exp[-ar^{b}+c]$ to fit the tails. We find that the exponent $b$ is sample-dependent and all samples have $b>1.0$, showing higher than exponential decay. Prior to this work, a large number of studies used the Lyapunov exponent as the core quantity~\cite{ShimaLocal,MouraDisorder,MouraDeloc,izrailev2012anomalous,kaya2007localization,zhao2010critical, PhysRevB.72.174207} and it is considered to be a localized phase for large $W$ and $\alpha$. However, the characteristics of the wave function for large $W$ and $\alpha$ are consistent with bound states for $\alpha=\infty$, suggesting that the bound effect of the potential is more dominant than the disorder effect. Thus, the so-called localized phase for large $W$ and $\alpha$ in the previous work is actually a bound phase. 

\begin{figure}[htbp]
	\includegraphics[width=1.0\columnwidth,height=0.8\columnwidth]{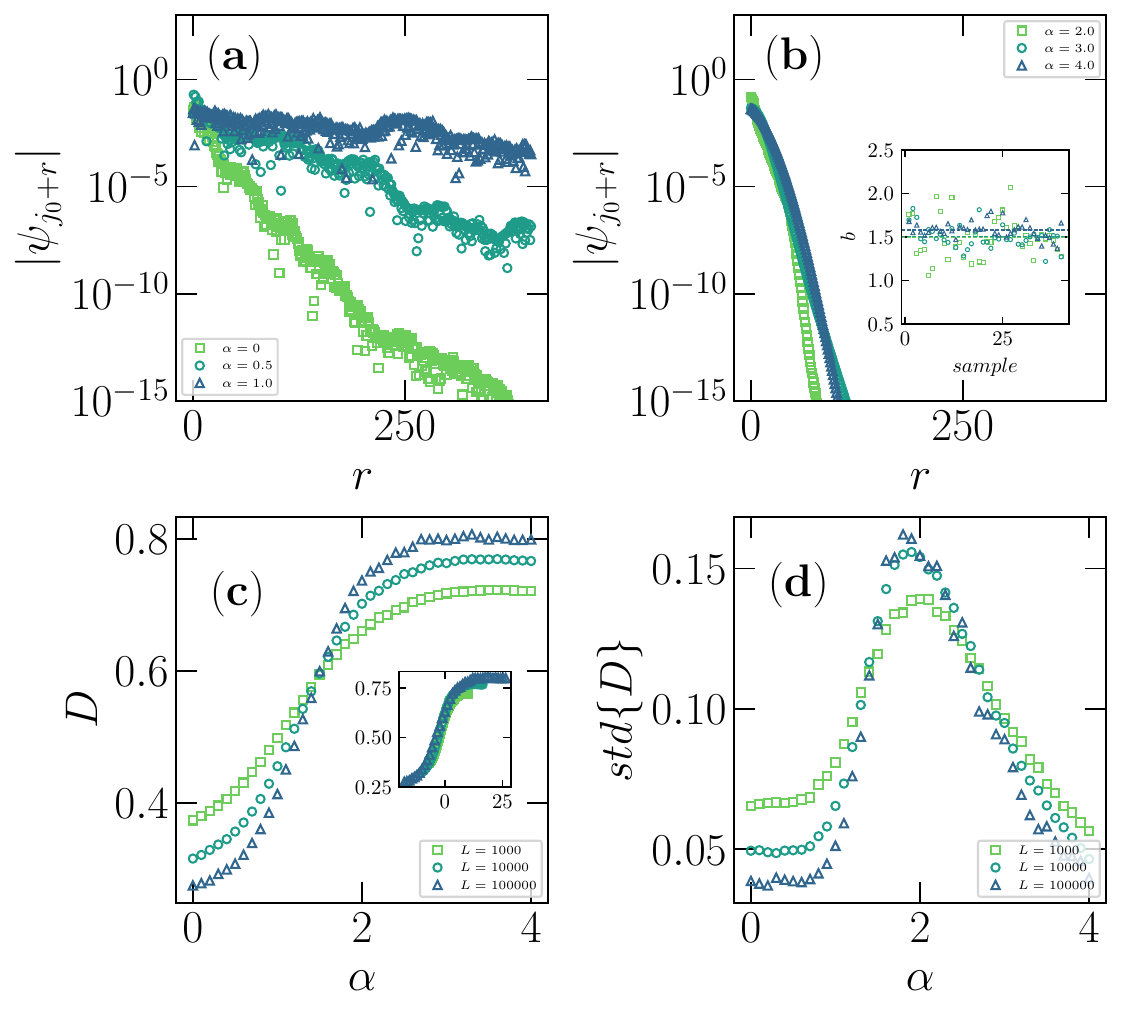}
	\vspace{-0.4cm}
	\caption{(a) and (b) The typical tails of Anderson localized states and bound states for different $\alpha$. (c) The fractal dimension as a function of $\alpha$ for different sizes. (d) The standard deviation of different samples of the fractal dimension as a function of $\alpha$ for different sizes. $W=6.0$. $L=10000$ in (a) and (b). The dashed lines in the inset of (b) represent the mean decay exponent. (c) and (d) 10000 samples are used. In the inset of (c), we scale the finite-size data using the ansatz $f[(\alpha-\alpha_c)L^{1/\nu}]$ and find $\nu \approx 4.8$, consistent with the Harris criterion $\nu > 2/d$~\cite{harris1974effect}.
	}
	\label{fig5}
\end{figure}

In addition to the transition from the extended phase to the bound phase, we also study the transition between the Anderson localized phase and the bound phase by fixing $W/t=6$ and adjusting $\alpha$. In Fig.~\ref{fig5} (a), we consider the case of weak $\alpha$, where the system is in the Anderson localized phase. As $\alpha$ increases, the correlation of colored noise strengthens, causing slower decay of the wave function and an increase in the Anderson localization length $\xi$. Compared with the Anderson localized states, the decay of the bound states for larger $\alpha$ is faster, as shown in Fig.~\ref{fig5} (b). This is because the bound states exhibit higher than exponential decay. In the inset of Fig.~\ref{fig5} (b), we show the decay exponents for different samples, and the mean decay exponents are marked by the dashed lines. The nearly equal decay exponents suggest that the decay of the wave function does not slow down significantly with increasing $\alpha$. To determine the transition point between the Anderson localized phase and the bound phase, we introduce the fractal dimension $D$. In Fig.~\ref{fig5} (c), fractal dimensions of different system sizes intersect at the transition point $\alpha\approx1.5$. For $\alpha>1.5$, the fractal dimension increases with the system size and one may expect $D\rightarrow1$ in the thermodynamic limit. On the contrary, as $L$ increases, the fractal dimension within the Anderson localized phase decreases and approaches $D=0$. In Fig.~\ref{fig5} (d), we study the standard deviation of the fractal dimension for different $\alpha$. Due to the mixing and coexistence of different phases near the transition in finite systems, large sample-to-sample fluctuations are expected near the transition point~\cite{Bera}. Fig.~\ref{fig5} (d) shows that the peaks occur at $\alpha\approx1.9$, which is slightly different from the transition point in Fig.~\ref{fig5} (c). This is attributed to the size effect, and they are expected to be consistent in the thermodynamic limit.

\subsection{the coexistence of different states in the spectrum}
\begin{figure}[htbp]
	\includegraphics[width=1.0\columnwidth,height=0.6\columnwidth]{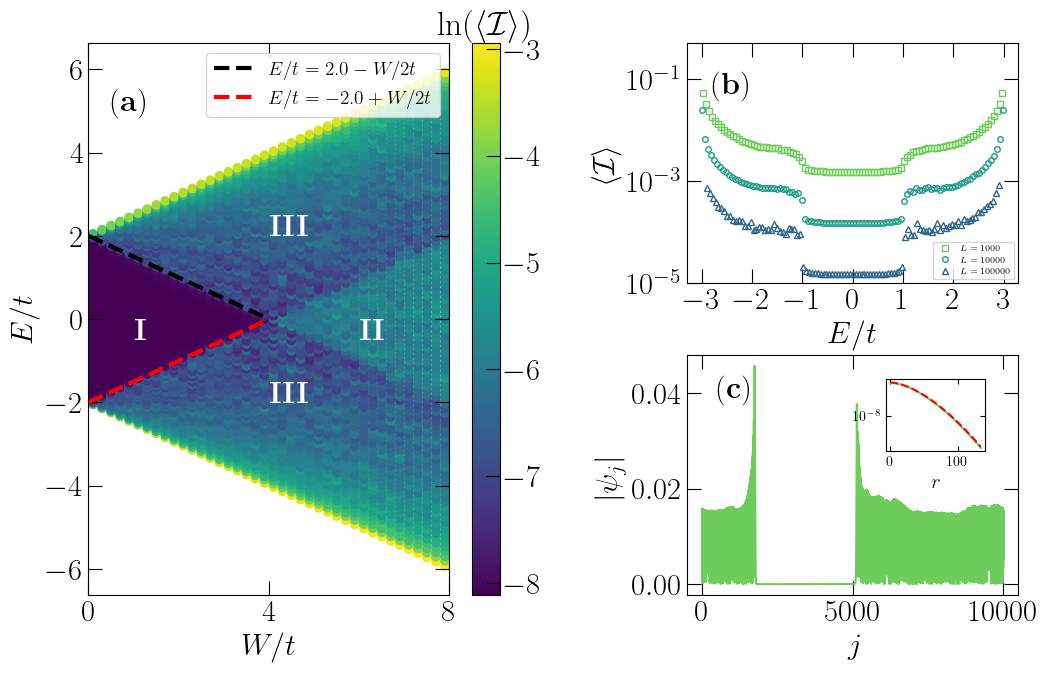}
	\vspace{-0.4cm}
	\caption{(a) The spectrum as a function of $W/t$ for $\alpha=3.0$. (b) $\langle \mathcal{I} \rangle$ as a function of energies for $\alpha=3.0$ and $W/t=2.0$. (c) The typical wave function for $E/t\approx2.0$ and one of its tails (left) is shown in the inset. $L=5000$ and 100 samples are used in (a). The red dotted line in the inset of (c) is the fitting function $|\psi_{1745+r}|=\exp[-0.0092r^{1.65}-2.876]$. 
	}
	\label{fig6}
\end{figure}

By fixing the energy $E_{target}\approx0$ above, we have studied the transition between different phases in the middle of the spectrum. To further understand the coexistence of different states in the spectrum, e.g., mobility edges~\cite{PhysRevLettIzrailev, MouraDisorder, PMoura,zhao2010critical,Russ134209,Petersen195443,diaz2005anomalous}, we scan the $E_{target}$ from the ground state energy to the highest state energy. In Fig.~\ref{fig6} (a), we show the spectrum as a function of $W/t$ for $\alpha=3.0$, which encodes three regions separated by critical energies marked by black and red dashed lines. These two critical energies can also be obtained by using energy conservation. The distribution of the wave function for large $\alpha$ is similar to that of the Wannier-Stark state induced by the linear potential~\cite{Wei134207}, whose distribution is mainly in the local energy interval $[E-2t, E+2t]$ (please see the Appendix~\ref{add2}), where $E$ is the eigenvalue. When this local energy interval is smaller than that of the potential amplitude $[-W/2, W/2]$, i.e.,  $E-2t\geq -W/2$ and $E+2t\leq W/2$, the atom becomes bound~\cite{Wei134207}. Thus critical energies satisfy $E_1-2t=-W/2$ and $E_2+2t=W/2$, that is, $E_1/t=2-W/2t$ and $E_2/t=-2+W/2t$. The properties of Regions I and II have been studied in Fig.~\ref{fig4} by taking $E_{target}\approx0$ as examples, here we focus on Region III. In Fig.~\ref{fig6} (b) we show $\langle \mathcal{I} \rangle$ as a function of energies for $W/t=2.0$, in which $\langle \mathcal{I} \rangle$ alters rapidly at the transition points $E_1/t=-1$ and $E_2/t=1$. An important point that needs to be highlighted is $\langle \mathcal{I} \rangle$ decreases with increasing size for $E/t<-1$ and $E/t>1$, corresponding to Region III in Fig.~\ref{fig6} (a). This suggests that the wave functions in Region III expand as the system size increases. By verifying the decay of the wave function, we confirm that it has a higher than exponential decay with the fitting exponent $b>1$. The typical wave function and its tail are shown in Fig.~\ref{fig6} (c). These characteristics collectively indicate that Region III is a bound phase. Thus, it can be concluded that a coexistence of extended states and bound states occurs for $W/t<4$ in Fig.~\ref{fig6} (a). In the Ref.~\cite{Nishino033105}, it is also described as the coexistence of extended states and gradon states. It should be noted that the key difference between our work and Ref.~\cite{Nishino033105} is that we analyze quantum phase coexistence from the perspective of the bound effect and derive the general expression for critical energies. Region III and Region II are both bound phases, but they have distinct boundaries in Fig.~\ref{fig6} (a). This is due to a difference in the distribution width of their wave functions, similar phenomena can also be observed in the Aubry-Andr\'{e}  model with a small incommensurate wave vector~\cite{wang2017almost}.

\begin{figure}[htbp]
	\includegraphics[width=1.0\columnwidth,height=0.6\columnwidth]{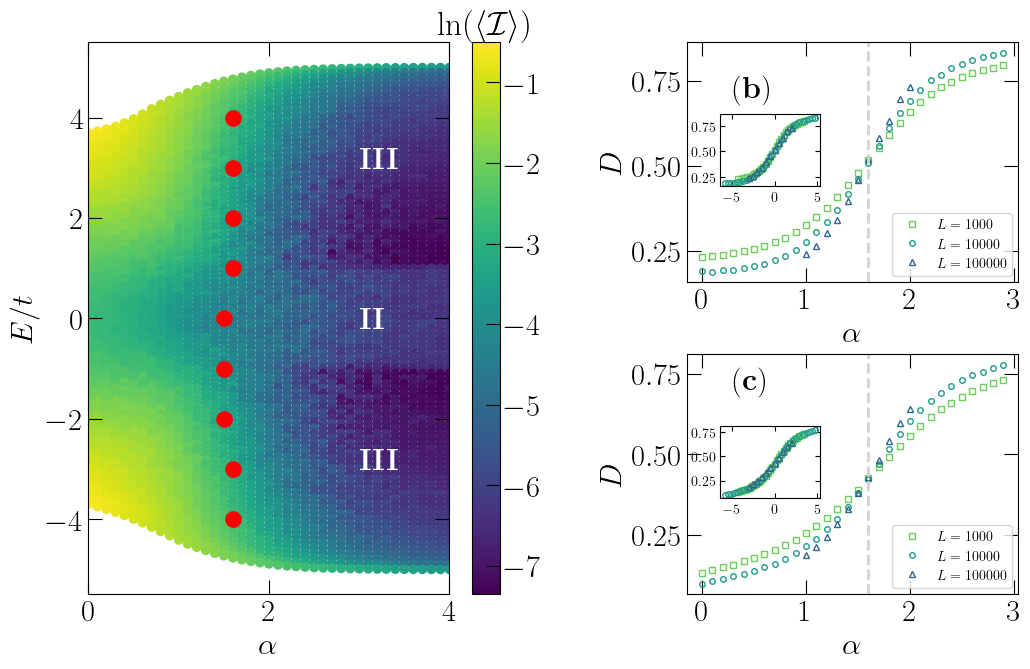}
	\vspace{-0.4cm}
	\caption{(a) The spectrum as a function of $\alpha$ for $W/t=6$. (b) and (c) The fractal dimension $D$ as a function of $\alpha$ for $E_{target}\approx2.0$ and $E_{target}\approx3.0$, respectively. $L=5000$ and 100 samples are used in (a) and the red markers indicate the transition points obtained by the scaling analysis of the fractal dimension $D$. Region III and Region II are bound phases, corresponding to those in Fig.~\ref{fig6} (a). 10000 samples are used and grey dotted lines mark the transition points at $\alpha_c=1.6$ in (b) and (c). In the insets of the right panel, we employ the ansatz $f[(\alpha-\alpha_c)L^{1/\nu}]$ to scale finite-size data and obtain $\nu\approx 7.1$.
	}
	\label{fig7}
\end{figure}

To further study the effect of $\alpha$ on the coexistence of different states, we fix $W/t=6$ and vary $\alpha$ to plot the spectrum in Fig.~\ref{fig7} (a). For $E/t\approx0$, the system undergoes a transition from the Anderson localized phase to the bound phase (Region II) and the transition point is at $\alpha_c\approx1.5$, as shown in Fig.~\ref{fig5} (c). In the other cases of high energies, marked by red circles in Fig.~\ref{fig7} (a), the transition points from the Anderson localized phase to the bound phase (Region II or Region III) are all close to $\alpha_c=1.5\sim1.6$, suggesting there is no coexistence of Anderson localized states and bound states in the spectrum for the highly excited states. Two typical examples for $E_{target}\approx2.0$ and $E_{target}\approx3.0$ are shown in Fig.~\ref{fig7} (b) and (c), in which the fractal dimensions for different sizes intersect at $\alpha_c\approx1.6$ for both cases. 
By employing the scaling form $f[(\alpha-\alpha_c)L^{1/\nu}]$~\cite{PhysRevBLuitz,LvPhysRevA}, which enables finite-size data to collapse onto a single curve, we obtain the exponent $\nu\approx 7.1$. This result is consistent with the Harris criterion $\nu > 2/d$~\cite{harris1974effect}, where $d$ represents the dimension of the system.
In Fig.~\ref{fig6}, we have shown that Region III and Region II have well-defined boundary energies at $E_1/t=2-W/2t$ and $E_2/t=-2+W/2t$, which are only determined by the potential amplitude. Thus, it shows $\alpha$-independent critical energies at $E_1/t=-1$ and $E_2/t=1$ between Region III and Region II in Fig.~\ref{fig7} (a).

\subsection{dynamical properties}

\begin{figure}[htbp]
	\includegraphics[width=1.0\columnwidth,height=0.6\columnwidth]{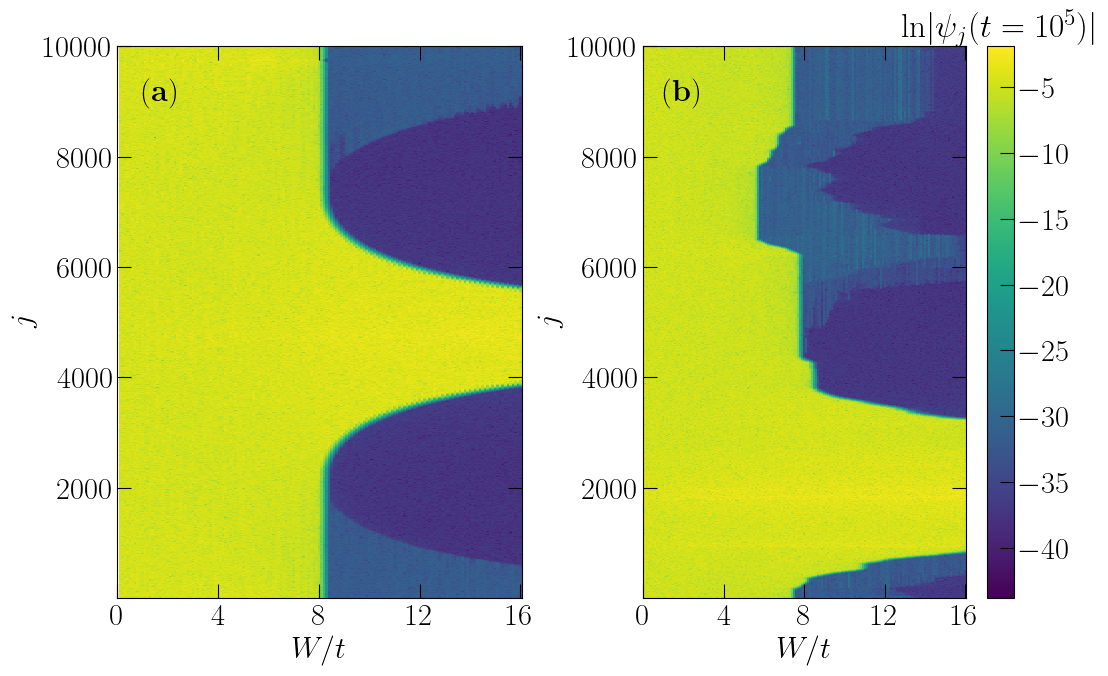}
	\vspace{-0.4cm}
	\caption{(a) and (b) The evolved wave functions as a function of $W/t$. In (a), $\alpha=\infty$, while in (b), $\alpha=3$. The horizontal axis is the potential amplitude $W/t$, and the vertical axis is the site index $j$. $L=10000$. The color represents the value of $\ln|\psi_j(t=10^5)|$. The atom is initially located at site with $\langle \psi(t=0)|\hat{H}|\psi(t=0)\rangle=0$, thus different potential configurations have different initial positions. In (a) and (b), the initial positions are at $j=4744$ and $j=1834$, respectively.
	}
	\label{fig8}
\end{figure}

In addition to the static (time-independent) analysis above, we also perform time evolution $\psi(t)=e^{-i\hat{H}t}\psi(t=0)$ to study the dynamical properties of the bound phase for large $\alpha$, where $\hat{H}$ is the Hamiltonian and $\psi(t=0)$ denotes the initial state. We trap the atom at a site where the initial energy satisfies  $\langle \psi(t=0)|\hat{H}|\psi(t=0)\rangle=0$. Since the adiabatic evolution keeps the energy unchanged, $\langle \psi(t)|\hat{H}|\psi(t)\rangle\equiv0$ for arbitrary time. We set the maximum evolution time to $T=100000$, at which the system has evolved to a steady state. We first consider the limiting case of $\alpha=\infty$ in Fig.~\ref{fig8} (a), in which the evolved wave function can be distributed throughout space for $W/t<8$, but cannot for $W/t>8$. Thus, the dynamic transition point from the extended phase to the bound phase is at $W/t=8$. In the Appendix~\ref{add1}, we show that all eigenstates are bound states for $W/t>4$, with the static transition point at $W/t=4$ for $\alpha=\infty$. This indicates that the dynamic transition point from the extended phase to the bound phase is inconsistent with the static transition point, in contrast to Anderson localization where the static and dynamic transition points are consistent~\cite{aubry1980analyticity,dominguez2019aubry}. The dynamic transition point is twice the static transition point, as determined by analyzing the distribution of eigenstates, as shown in Ref.~\cite{Wei023314}. Furthermore, it can be observed that even with a large potential amplitude, the evolved wave function exhibits a wide distribution. Typically, the evolved wave function spans from $j\approx3900$ to $j\approx5700$ for $W/t=16$. This can be attributed to the bound effect, which requires an extremely large amplitude to confine the atom to a single site. In Fig.~\ref{fig8} (b), we show the evolved wave function as a function of $W/t$ for $\alpha=3$, in which the dynamical transition point is at $W/t=5.6$. This differs from the static transition point at $W/t=4$ in Fig.~\ref{fig4}, indicating that the transition points between static and dynamical cases are inconsistent for large $\alpha$, as in the limiting case of $\alpha=\infty$. It also needs to be pointed out that the static (time-independent) analysis in Fig.~\ref{fig3} and Fig.~\ref{fig4} shows that the static transition points for $\alpha=3$ and $\alpha=\infty$ are consistent. However, in the dynamic case, we find that the transition points for $\alpha=3$ vary between samples and may differ from those for $\alpha=\infty$, as shown in Fig.~\ref{fig8}. In Appendix~\ref{add4}, we also provide the dynamical phase diagram for comparison with the static phase diagram.

\section{conclusion}
In summary, we have studied an atom in a one-dimensional system subjected to colored noise. Different from previous works that used the Lyapunov exponent as the core quantity, we mainly focus on a more systematic description of the transition based on the wave functions in the present work. By studying the spatial distribution, tail decay of wave functions, inverse participation rate, and fractal dimension, we find that a bound phase exists for large $W$ and $\alpha$. The essential reason for the formation of this phase is that the correlation of the potential strengthens as $\alpha$ increases, resulting in the bound effect of the potential being more dominant than disorder. By studying the properties of quantum states in the spectrum, we confirm that extended and bound states coexist with well-defined critical energies for large $\alpha$. In contrast, when the system undergoes transitions from the Anderson localized phase to the bound phase as $\alpha$ increases, we do not observe the obvious coexistence of different quantum states for the highly excited states in the spectrum. After introducing adiabatic evolution, we find that the dynamic transition point induced by the bound effect is inconsistent with the static transition point. This is different from the disorder-induced Anderson localization transition, where the dynamic and static transition points coincide.

\begin{acknowledgments}
We acknowledge support from the Natural Science Foundation of China (Grants No. 12404322, No. 12074340, and No. 12474492),
the Zhejiang Provincial Natural Science Foundation of China under Grant No.LQ24A040004,
the Natural Science Foundation of Jiangsu Province (Grant No. BK20200737),
and the Science Foundation of Zhejiang Sci-Tech University (Grant Nos. 23062152-Y and 20062098-Y). We acknowledge
 enlightening discussions with Niaz Ali Khan.
\end{acknowledgments}

\bibliography{reference}

\clearpage

\onecolumngrid

\appendix
\begin{appendices}
	
\vspace{0.3cm}
	
\twocolumngrid
	
\beginsupplement

\section{Results of regulating the variance of the potential}~\label{add3}
\begin{figure}[htbp]
	\includegraphics[width=1.0\columnwidth,height=0.8\columnwidth]{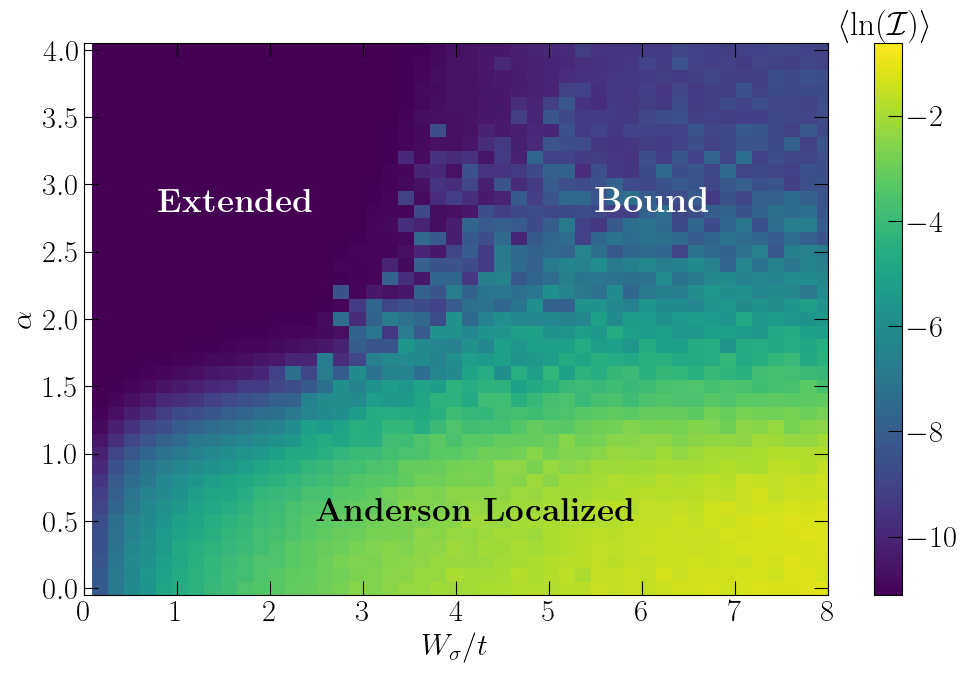}
	\vspace{-0.4cm}
	\caption{ Phase diagram in the middle of the spectrum $E_{target}\approx0$. Applying variance normalization through the formula $\varepsilon_j = (\tilde{\varepsilon}_j -\langle \tilde{\varepsilon}_j \rangle)/std\{\tilde{\varepsilon}_j\}$, parameters are the same as those in Fig.~\ref{fig2}.
	}
	\label{figadd3}
\end{figure}

In the main text, we normalize the potential amplitude using $\varepsilon_j=-1/2+(\tilde{\varepsilon}_j-\min\{\tilde{\varepsilon}_j\})/(\max\{\tilde{\varepsilon}_j\}-\min\{\tilde{\varepsilon}_j\})$. Here, we apply variance normalization through the formula $\varepsilon_j = (\tilde{\varepsilon}_j -\langle \tilde{\varepsilon}_j \rangle)/std\{\tilde{\varepsilon}_j\}$, where $\langle \cdot \rangle$ denotes the mean value and $std\{ \cdot \}$ represents the standard deviation. The variance normalization formula ensures that  $\langle \varepsilon_j \rangle = 0$ and $\sqrt{\langle {\varepsilon}^2_j \rangle -\langle {\varepsilon}_j \rangle^2 }=1$. To distinguish the amplitude $W$ in the main text, we set the potential amplitude after variance adjustment to $W_{\sigma}/\sqrt{12}$, following the method in Ref.~\cite{kaya2007localization}. Fig.~\ref{figadd3} shows that the transition points from the extended phase to the bound phase are $\alpha$-denpendent for large $\alpha$, which is consistent with Ref.~\cite{kaya2007localization}. After carefully checking the transition points of individual samples, we find that the transition points of different samples are different, consistent with  Ref.~\cite{Nishino033105}. The underlying reason of this phenomenon is that the transition point depends on the amplitude, and variance normalization does not ensure consistent amplitude across samples.

\begin{figure}[htbp]
	\includegraphics[width=1.0\columnwidth,height=0.8\columnwidth]{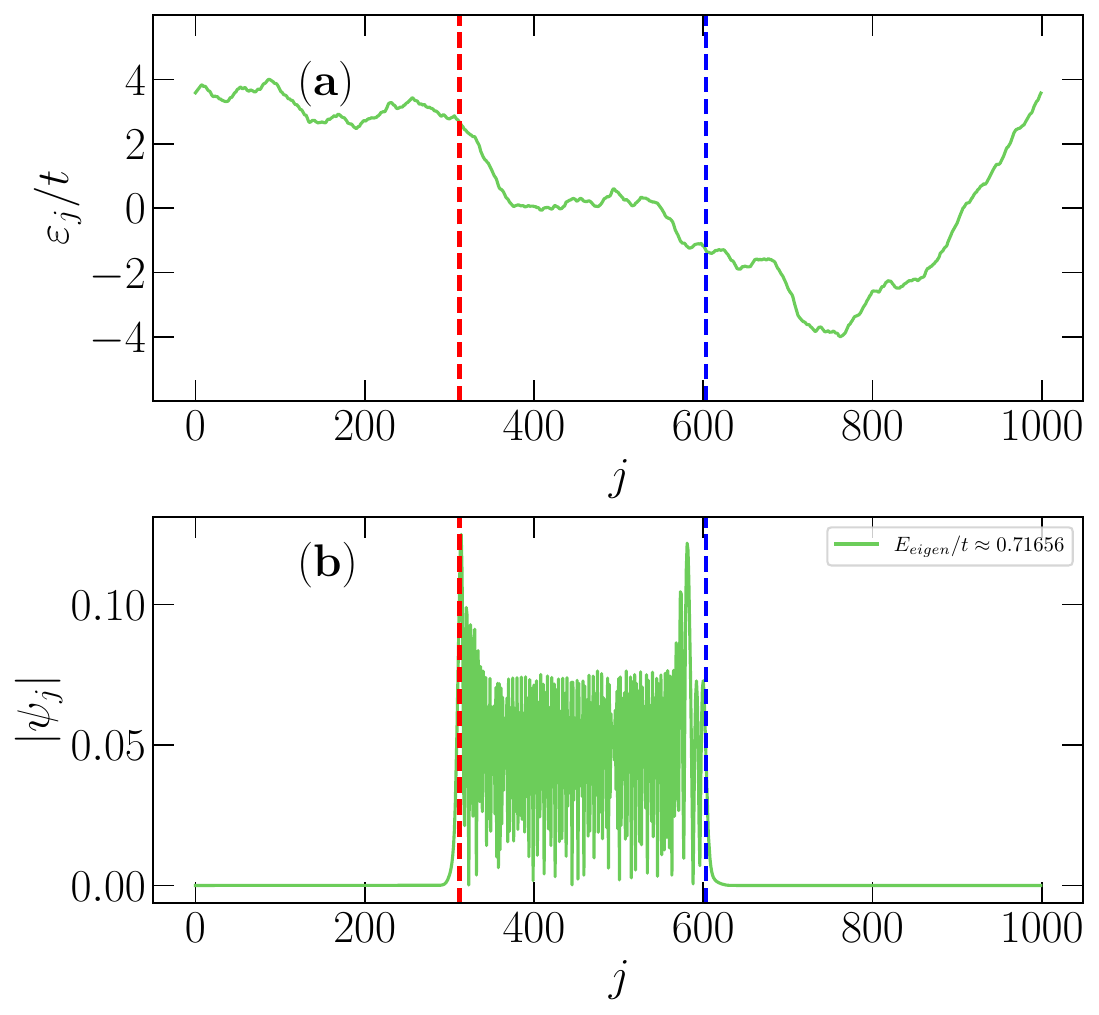}
	\vspace{-0.4cm}
	\caption{(a) One example of colored noise for $\alpha=3$. (b) The eigenstate at the eigenvalue $E_{eigen}/t\approx0.71656$.  $L=1000$ and the colored noise is controlled to [-4, 4] with the amplitude $W/t=8$.  
	}
	\label{figadd1}
\end{figure}

\section{Distribution of bound states}~\label{add2}
In Fig.~\ref{figadd1} (a), we show an example of colored noise for $\alpha=3$, whose interval is adjusted to $[-4, 4]$ by setting $W/t=8$. Through exact diagonalization, we obtain all the eigenvalues and eigenstates. Then, we arbitrarily extract one eigenstate and display it in Fig.~\ref{figadd1} (b). The corresponding eigenvalue is $E_{eigen}/t\approx0.71656$. According to Ref.~\cite{Wei134207}, bound states caused by the binding effect are mainly distributed within the local energy interval $\varepsilon_j/t\in[E_{eigen}/t-2, E_{eigen}/t+2]$. In Fig.~\ref{figadd1} (a), the red and blue dotted lines indicate the sites at $j=312$ and $j=603$, respectively. The local energies corresponding to these two sites are approximately $\varepsilon_j/t=E_{eigen}/t-2=-1.28344$ and $\varepsilon_j/t=E_{eigen}/t+2=2.71656$. In Fig.~\ref{figadd1} (b), the red and blue dotted lines are the same as those in Fig.~\ref{figadd1} (a). It can be found that the eigenstate is mainly distributed between the two dotted lines, indicating that the conclusion regarding the distribution of bound states in Ref.~\cite{Wei134207} is also applicable to our work for large $\alpha$.

\section{The spectrum for $\alpha=\infty$}~\label{add1}
\begin{figure}[htbp]
	\includegraphics[width=1.0\columnwidth,height=0.8\columnwidth]{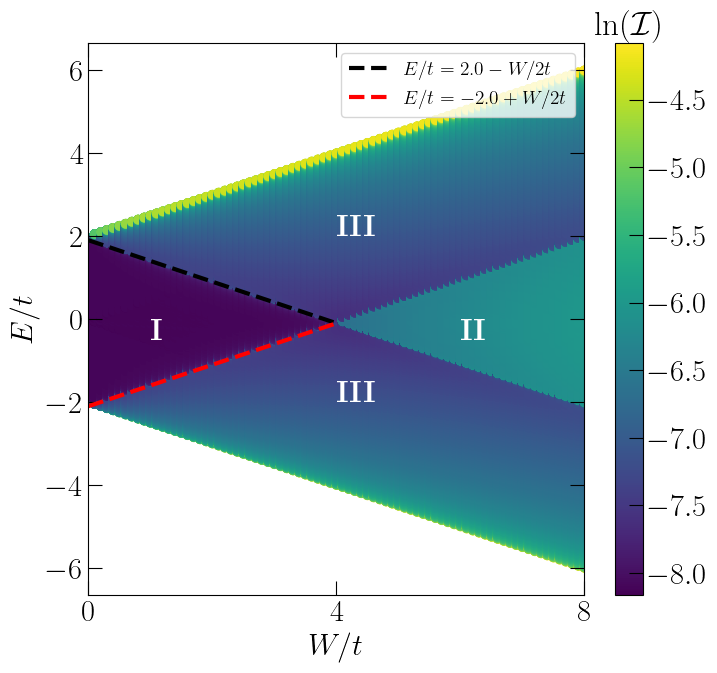}
	\vspace{-0.4cm}
	\caption{The spectrum as a function of $W/t$ for $\alpha=\infty$. $\tilde{\varepsilon}_j = \cos(2\pi j/L+\phi_{1})$ for $\alpha=\infty$ and $\phi_1=\sqrt{3}$. The phase labels are the same as those in  Fig.~\ref{fig6} (a) in the main text. 
	}
	\label{figadd2}
\end{figure}

In Fig.~\ref{figadd2}, we show the spectrum as a function of $W/t$ for $\alpha=\infty$, where the composition of the phase diagram and the critical energies between different phases are consistent with those in Fig.~\ref{fig6} (a). The consistency of both phase diagrams further confirms the conclusion that the bound effect of the potential is more dominant than the disorder effect for large $\alpha$. Furthermore,  it can be found that extended states (Region I) and bound states (Region III) coexist for $W/t<4$, whereas all states are bound states for $W/t>4$. Note that Region II and Region III are both bound phases.

\begin{figure}[htbp]
	\includegraphics[width=1.0\columnwidth,height=0.8\columnwidth]{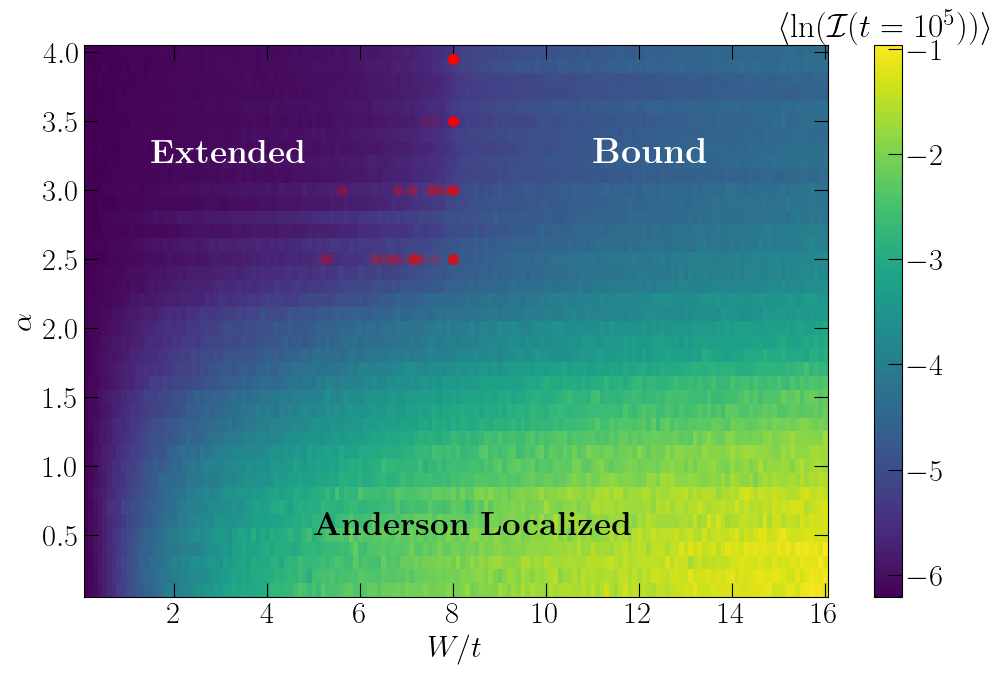}
	\vspace{-0.4cm}
	\caption{Dynamical phase diagram after long-time evolution. The atom is initially located at site with $E=\langle \psi(t=0)|\hat{H}|\psi(t=0)\rangle=0$. The colors represent the values of $\ln(\langle\mathcal{I}(t=10^5)\rangle)$, where $\mathcal{I}(t)=\sum_j|\psi_j(t=10^5)|^{4}$. $L=1000$. The red circles mark the transition points across 20 samples, with darker shades indicating higher frequencies of occurrence.
	}
	\label{figadd4}
\end{figure}

\section{Dynamical phase diagram}~\label{add4}
In Fig.~\ref{fig2}, we present the static phase diagram in the middle of the spectrum $E_{target}\approx0$. Here we plot the dynamical phase diagram by calculating the time-dependent inverse participation ratio $\mathcal{I}(t)=\sum_j|\psi_j(t)|^{4}$ and 20 examples are used in Fig.~\ref{figadd4}. At time $t=0$, we trap the atom at a site that satisfies the condition $E=\langle \psi(t=0)|\hat{H}|\psi(t=0)\rangle=0$ . The maximum evolution time is $T=10^5$, ensuring the system reaches a steady state after long-time evolution. Compared to Fig.~\ref{fig2}, the dynamical phase diagram reveals that the extended phase occupies a larger region of parameter space. In the static diagram, the transitions from the extended phase to the bound phase occur at $W/t=4$ in Fig.~\ref{fig2}, whereas in the dynamical diagram they shift to $W/t>4$. For $\alpha=4.0$, the transition points of 20 examples are all at $W/t=8$, which is the same as the case of $\alpha=\infty$. As $\alpha$ decreases, the transition points (red markers in Fig.~\ref{figadd4}) become sample-dependent and deviate from $W/t=8$, indicating the stronger disorder effect competing with the bound effect as $\alpha$ decreases.

\end{appendices}

\end{document}